\documentclass[aps,pra,floatfix,twocolumn,showpacs,amsmath,amssymb]{revtex4}
\input{epsf}
\usepackage{graphicx}
\usepackage{array}
\usepackage{bigstrut}
\usepackage{longtable}
\usepackage{rotating,booktabs}
\usepackage{booktabs,threeparttable}
\usepackage{diagbox}
\usepackage[mode=text]{siunitx}
\usepackage{bm}
\usepackage{lipsum}
\usepackage{warpcol}
\usepackage{color}
\usepackage{amsthm,amsmath,amssymb}
\usepackage{mathrsfs}
\usepackage{appendix}
\usepackage{multirow}
\usepackage{threeparttable}

\begin{document}

\title{Application of the correlated B-spline basis functions to the leading relativistic and QED corrections of helium}

\author{Hao Fang$^{1, 2}$, Yong-Hui Zhang$^{1}$, Pei-Pei Zhang$^1$, and Ting-Yun Shi$^{1, \dag}$~\footnotetext{$\dag$Email Address: tyshi@wipm.ac.cn}}

\affiliation {$^1$State Key Laboratory of Magnetic Resonance and Atomic and Molecular Physics, Wuhan Institute of Physics and
Mathematics, Innovation Academy for Precision Measurement Science and Technology, Chinese Academy of Sciences, Wuhan 430071, People's Republic of China}

\affiliation {$^2$University of Chinese Academy of Sciences, Beijing 100049, People's Republic of China}

\date{\today}

\begin{abstract}
    B-spline functions have been widely used in computational atomic physics. Different from the traditional B-spline basis (a simple product of two B-splines), the recently developed correlated B-spline basis functions(C-BSBF), in which the interelectronic coordinate $r_{12}$ is included explicitly, have greatly improved the computational accuracy of polarizability [S. J. Yang \textit{et al}., Phys. Rev. A \textbf{95}, 062505 (2017)] and bethe logarithm [ S. J. Yang \textit{et al}., Phys. Rev. A \textbf{100}, 042509 (2019)] for singlet states of helium. Here, we report the extension of the C-BSBF to the leading relativistic and QED correction calculations for energy levels of the $1\,^1S$, $2\,^1S$, $2\,^3S$, and $3\,^3S$ states of helium. The relativistic kinetic term $p_{1}^{4}$, contact potential $\delta^{3}(r_{1})$, $\delta^{3}(r_{12})$ and Araki-Sucher correction $\langle 1/r_{12}^{3} \rangle$ are calculated by using the global operator method, in which $r_{12}^n$ and $r_{12}^n\ln r_{12}$ involved are calculated with the generalization of Laplace's expansions. The obtained values for the ground state are $\delta E_{rel}/\alpha ^{2}=-$1.951 754 7(2) and $\delta E_{QED}/\alpha^{3}=$57.288 165(2), consistent with previous results, which opens the possibility of calculating higher-order relativistic and QED effects using the C-BSBF.

%Our results are consistent with the best results in the literature. The correlated B-spline basis set is a more advantageous candidate basis set, which has the ability to calculate higher singularity operators.
\end{abstract}

\maketitle

\section{introduction}

The measured precision of helium atomic spectroscopy has approached the part-per-trillion level~\cite{rengelink18, zheng17b}, which allows the test of quantum electrodynamics (QED) and the determination of the fine-structure constant $\alpha$ and the nuclear charge radius~\cite{clausen21, kato18, zheng17a, rengelink18, zheng17b, pastor12, shiner95} by combination with the high-accuracy atomic structure calculations~\cite{patkos21, patkos16, patkos17, yerokhin10}. In addition, from the theoretical point of view, as the simplest many-electron system, traditionally helium is an ideal testing ground for different methods of the description of atomic structure.

It is known that finite basis set variational calculations are the most powerful tool for solving the Coulomb three-body bound-state problem exactly, such as helium, in which their basis functions included explicitly the interelectron separation are particularly important. For example, using the explicitly correlated exponential basis with nonlinear parameters, Pachucki et al. have accomplished complete $\alpha^7$m Lamb shift of helium triplet states, which improved the theoretical accuracy of ionization energies by more than an order of magnitude~\cite{patkos21}. Hylleraas variation technique is employed to finish the calculations of the hyperfine structure of the $2\,^3P_J$ state in $^7$Li$^+$ up to order m$\alpha^6$, which has improved previous calculations by one order of magnitude~\cite{qi20}. However, in order to get rid of loss of stability when the number of basis functions increases, these high-precision calculations must be supplemented by applying multiprecision package as well as variational optimized nonlinear parameters.

B-splines have the property of being ‘complete enough’ and linear independence even for a large basis, which has been widely used in computational atomic physics~\cite{Johnson86, bachau01, arimondo08, Bian08b, Zhang12, tang13, Hu14, Zhang16, Tang17, Zhang20}. With the development of high-resolution atomic spectroscopy, calculations of highly accurate energies are required. However, high-accuracy computational results are difficult to achieve with the traditional B-spline basis functions, for systems with strong electron correlations. For example, Lin et al.\cite{Lin12} gave a nonrelativistic ground state energy of $-$2.903 582 0 for the helium by using the B-spline basis, which had four accurate figures at the cost of a large number of the conﬁgurations. Also the relativistic energy for the $2\,^1S_0$ state of helium given by using the partial wave $\ell_{max}$=15 was only with six accurate figures~\cite{wu18}. So it is necessary to introduce the interelectronic coordinate into the traditional B-spline basis.

Recently, Tang et al. developed a method to calculate the Bethe logarithm, the dominant part of QED, of the hydrogen atom using the B-spline basis set~\cite{tang13}, which not only can calculate low-lying states with high precision using relatively small basis sets, but also can calculate highly-excited Rydberg states. Then Zhang et al. extended it to calculate the Bethe logarithms for the $S$ state of the helium atom~\cite{Zhang20}, in which the Bethe logarithms for the triplet state with weak electron correlation can be reached with five to eight accurate figures, but the precision is limited for the single state as the electron correlation effect is not included in the basis set. Therefore, Yang et al. have developed the explicitly correlated B-spline basis method and successfully applied it to the calculation of energy levels, static dipole polarizabilities~\cite{yang17}, and Bethe logarithms~\cite{yang19} for the singlet states of the helium atom. The nonrelativistic ground state energy has reached $-$2.903 724 377 1(2)~\cite{yang17}, which is six orders of magnitude better than the result of Lin et al.~\cite{Lin12}. Moreover, they have been able to obtain static dipole polarizabilities with a relative error of 10$^{-9}$ and Bethe logarithms with a relative error of 10$^{-7}$, respectively, which shows that the correlated B-spline basis functions(C-BSBF) can describe well the electronic correlation of the singlet states and effectively improve the numerical convergence rates.

This work will employ the C-BSBF to evaluate the leading relativistic and QED corrections to energy levels of the helium atom. The global operator method will be used to improve the numerical convergence for the relativistic kinetic term $p_{1}^{4}$, contact potential $\delta^{3}(r_{1})$, $\delta^{3}(r_{12})$ and Araki-Sucher correction $\langle 1/r_{12}^{3} \rangle$, which will expand the scope of using of the C-BSBF and present a manifestation that the C-BSBF can be effectively applied to numerical calculations of the expectation values of singular operators as well.

This paper is organized as follows. The theoretical formulas and methods used in our calculations are presented in section II. In section III we calculate the leading relativistic and QED corrections to energy levels for the $1\,^1S$, $2\,^1S$, $2\,^3S$ and $3\,^3S$ states of helium. Comparisons with results of available literature are made as well. Conclusions are given in section IV. Atomic units (a.u.) are used throughout this paper.

\section{theory and method}

\subsection{Correlated B-spline basis functions(C-BSBF)}

The nonrelativistic Hamiltonian for a two-electron atom with an infinite mass nucleus has the form of
\begin{eqnarray}
H=\sum\limits_{i=1}^2\left(\dfrac{{\bf{p}}_{i}^{2}}{2}-\dfrac{Z}{r_i}\right)+\frac{1}{r_{12}}\,,\label{e1}
\end{eqnarray}
where ${\bf{p}}_{i}=-i\nabla_i$ is the momentum operator of the $i$th electron, $r_i$ is the coordinate of the $i$th electron to the atomic nucleus, $r_{12}$ is the interelectronic coordinate, and the nuclear charge $Z=2$ for the helium atom.

The two-electron wave function is expanded by the following C-BSBF in which the interelectronic coordinate $r_{12}$ is included explicitly,
\begin{eqnarray}
\phi_{ij, c, \ell_1\ell_2}=\mathcal{A}\left[ r_{12}^{c} B_{i}^{k}\left(r_{1}\right) B_{j}^{k}\left(r_{2}\right) \mathcal{Y}_{\ell_{1} \ell_{2}}^{L M}\left(\hat{\bf{r}}_{1}, \hat{\bf{r}}_{2}\right)\right]\,,
\end{eqnarray}
where the operator $\mathcal{A}$ ensures the antisymmetry of the basis function with respect to the exchange of the two electrons, $B_{i}^{k}(r)$ is the $i$th of $N$ B-spline functions with the order of $k$ and constrained to a spherical cavity~\cite{bachau01}, $c$ is the power of the $r_{12}$ coordinate, and the coupled spherical harmonic function is given by
\begin{eqnarray}
    \begin{aligned}
\mathcal{Y}_{\ell_{1} \ell_{2}}^{L M}\left(\hat{\bf{r}}_{1}, \hat{\bf{r}}_{2}\right)=\sum\limits_{m_{1} m_{2}}&\left\langle \ell_{1} \ell_{2} m_{1} m_{2} \mid L M\right\rangle \\
&\times Y_{\ell_{1} m_{1}}\left(\hat{\bf{r}}_{1}\right) Y_{\ell_{2} m_{2}}\left(\hat{\bf{r}}_{2}\right)\,,
    \end{aligned}
\end{eqnarray}
with $\left\langle \ell_{1} \ell_{2} m_{1} m_{2} \mid L M\right\rangle$ being the Clebsch-Gordan coefficient. In the present calculations, the cavity radius of $R_0$ is chosen appropriately, the $r_{12}$ power $c$ is restricted to be 0 or 1 without making integral evaluations overly complicated, and the orbital angular momentum $\ell_1$ and $\ell_2$ are less than the maximum partial wave $\ell_{max}$.

\subsection{Leading relativistic and QED corrections}

The leading relativistic correction to the nonrelativistic energy of the two-electron atom is given by the expectation value of the Breit-Pauli Hamiltonian with the nonrelativistic wave function $\psi$,
\begin{eqnarray}
\delta E_{rel}=\langle\psi|H_{BP}|\psi\rangle\,,
\end{eqnarray}
where
\begin{eqnarray}
H_{BP}=\alpha^{2}\left\{-\frac{1}{8}\left(p_{1}^{4}+p_{2}^{4}\right)+\pi \delta^{3}\left(r_{12}\right)+\dfrac{Z\pi}{2}\left[\delta^{3}\left(r_{1}\right)\right.\right.\nonumber\\ \left.+\delta^{3}\left(r_{2}\right)\right]\left.-\frac{1}{2 r_{12}} \left({\bf{p}}_1\cdot{\bf{p}}_2+\frac{{\bf{r}}_{12}\cdot({\bf{r}}_{12}\cdot{\bf{p}}_{1}){\bf{p}}_2}{r_{12}^{2}}\right)\right\}\,,\label{e5}
\end{eqnarray}
for $S$-state~\cite{stone61,drake88,yerokhin10}, where $\alpha=$7.297 352 569 3(11)$\times 10^{-3}$~\cite{tiesinga21} is the fine structure constant, $\delta^{3}(r_{12})$, $\delta^{3}(r_{1})$, and $\delta^{3}(r_{2})$ represent the Dirac delta functions.
The last term of Eq.~(\ref{e5}) is a retardation term, since this correction is due to the retardation of the electromagnetic
field produced by an electron~\cite{bethe12}, and $-\left[{\bf{p}}_1\cdot{\bf{p}}_2+{\bf{r}}_{12}\cdot({\bf{r}}_{12}\cdot{\bf{p}}_{1}){\bf{p}}_2/r_{12}^{2}\right]/2r_{12}$ is labelled as $H_2$.

The leading QED correction can be expressed as an expectation value of the following effective operators~\cite{yerokhin10, araki57, sucher58},
\begin{eqnarray}
\delta E_{QED}=\alpha^3\left\{\dfrac{4 Z}{3}\left[\frac{19}{30}-2\ln \alpha-\ln k_0\right]\langle\psi|\delta^3(r_1)\right. \nonumber \\
\left.+\delta^3(r_2)|\psi\rangle+\left[\dfrac{164}{15}+\dfrac{14}{3} \ln \alpha\right]\langle\psi|\delta^3(r_{12})|\psi\rangle \right.  \nonumber \\
\left.-\dfrac{7}{6\pi}\langle\psi|r_{12}^{-3}|\psi\rangle\right\}\,.\label{e6}
\end{eqnarray}
\begin{table}[!htbp]
    \caption{\label{t0} Bethe logarithm for the $1\,^1S$, $2\,^1S$, $2\,^3S$ and $3\,^3S$ states of helium.}
    \begin{ruledtabular}
    \begin{tabular}{clll}
    \multicolumn{1}{c}{State} &\multicolumn{1}{c}{Zhang\cite{Zhang20} and Yang\cite{yang19} }  &\multicolumn{1}{c}{Korobov\cite{korobov19}}\\ \hline
    $1^{1}S$                 &4.370 160 22(5)             &4.370 160 223 070 3(3) \\
    $2^{1}S$                 &4.366 412 71(1)             &4.366 412 726 417(1) \\
    $2^{3}S$                 &4.364 036 7(2)              &4.364 036 820 476(1) \\
    $3^{3}S$                 &4.368 666 7(1)              &4.368 666 996 159(2) 
    \end{tabular}
    \end{ruledtabular}
    \end{table}
Here $\ln k_0$ is the Bethe logarithm, and the last term in Eq.~(\ref{e6}) is usually called Araki-Sucher correction~\cite{araki57, stanke17, frolov05}, and the expectation of $\langle\psi|r_{12}^{-3}|\psi\rangle$ is defined as %
\begin{eqnarray}
    \begin{aligned}
\langle\psi|r_{12}^{-3}|\psi\rangle=\lim _{a \rightarrow 0} & \langle r_{12}^{-3}\Theta(r_{12}-a) \\
&+4 \pi(\gamma+\ln a) \delta^3(r_{12})\rangle\,,
    \end{aligned}
\end{eqnarray}
where $\Theta(x)$ and $\gamma$ are the step function and the Euler constant, respectively. Compared with the relativistic correction, the more difficult to calculate in the leading QED correction are Bethe logarithm and Araki-sucher correction. The Bethe logarithms for the $1\,^1S$, $2\,^1S$, $2\,^3S$ and $3\,^3S$ state of the helium atom are summarized in Table~\ref{t0} calculated by Zhang et al.\cite{Zhang20} using traditional B-spline function and Yang et al.\cite{yang19} using the C-BSBF, respectively, which based on the Drake-Goldman’s method. The Korobov's results listed in the last column of Table~\ref{t0} based on the integral representation method of Schwartz are the benchmarks. The value of the Bethe logarithms from Zhang et al. and Yang et al. are used in this work, which will achieve the complete calculation of the leading relativistic and QED correction using the B-spline function.

Drachman proposed the global operator method to evaluate the two-particle contact potential $\delta^3(r_{1})$ and $\delta^3(r_{12})$, which achieved significant improvements over the direct evaluations~\cite{drachman81}. We employ the equivalent form containing global operators Drachman given to calculate the expectation value of $\delta^3(r_{1})$ and $\delta^3(r_{12})$,
\begin{eqnarray}
    \begin{aligned}
        4\pi\left\langle\psi\left|\delta^{3}(r_{i})\right|\psi\right\rangle =&4\langle\psi|r_{i}^{-1}(E_{\psi}-V)|\psi\rangle \\
        &-2\sum\limits_{s=1}^2\langle\nabla_s\psi|r_{i}^{-1}|\nabla_s\psi\rangle\,,
            \end{aligned} 
\end{eqnarray}
\begin{eqnarray}
    \begin{aligned}
4\pi\left\langle\psi\left|\delta^{3}(r_{12})\right|\psi\right\rangle =&2\langle\psi|r_{12}^{-1}(E_{\psi}-V)|\psi\rangle \\
&-\sum\limits_{s=1}^2\langle\nabla_s\psi|r_{12}^{-1}|\nabla_s\psi\rangle\,,
    \end{aligned}
\end{eqnarray}
where $E_{\psi}$ is the corresponding eigenvalue of the two-electron wave function $\psi$, and $V=-{Z}/{r_{1}}-{Z}/{r_{2}}+{1}/{r_{12}}$.
It will result in a slow convergence for the kinetic term $p_1^4+p_2^4$ in the relativistic correction if we calculate its expectation value directly in the C-BSBF. Pachucki and Komasa also used a similar way to transform both the kinetic term and the Araki-Sucher correction to much more regular forms and obtained much better numerical convergence on that account~\cite{pachucki04}. In the present calculations, as Pachucki and Komasa have done, we use the following expression to evaluate $\langle p_1^4+p_2^4\rangle$,
\begin{eqnarray}
\sum_{i=1}^{2}\left\langle\psi \left| p_{i}^{4} \right|\psi\right\rangle=4\left\langle  \psi \left| (E_{\psi}-V)^2\right| \psi
					 \right\rangle -2\left\langle \nabla _{1}^{2} \psi|\nabla_{2}^{2}\psi\right\rangle\,. \label{e8}
\end{eqnarray}
The integration of $\langle\psi|r_{12}^{-2}|\psi\rangle$ will be involved in Eq.~(\ref{e8}), and it is also evaluated to be as following by using the global operator method,
\begin{eqnarray}
\begin{aligned}
\left\langle\psi\left|{r_{12}^{-2}}\right|\psi\right\rangle =&2\langle\psi|\ln r_{12}(V-E_{\psi})|\psi\rangle \\
&+\sum\limits_{i=1}^2\langle\nabla_i\psi|\ln r_{12}|\nabla_i\psi\rangle\,,\label{e9}
\end{aligned}
\end{eqnarray}
since we find that $\nabla^{2}_{1} \ln r_{12}=\nabla^{2}_{2} \ln r_{12}=r_{12}^{-2}$. The complete expansion of Eq.~(\ref{e8}) is written as
    \begin{eqnarray}
        \begin{aligned}
    \sum_{i=1}^{2}\left\langle\psi \left| p_{i}^{4} \right|\psi\right\rangle=4E_{\psi}^{2}-8E_{\psi}\left\langle \psi \left|-\frac{2Z}{r_{1}}+\frac{1}{r_{12}}\right| \psi\right\rangle &\\
    +4 \left\langle  \psi \left| \frac{2Z^{2}}{r_{1}^{2}}-\frac{2Z^{2}}{r_{1}r_{2}}-\frac{2Z}{r_{1}r_{12}}+\frac{1}{r_{12}^{2}}\right| \psi\right\rangle &\\
    -2\left\langle \nabla _{1}^{2} \psi|\nabla_{2}^{2}\psi\right\rangle  &\,. \label{P^{4}}
        \end{aligned}
    \end{eqnarray}
The Araki-Sucher correction is converted to the regular form as well so as to facilitate the present numerical evaluations,
\begin{eqnarray}
    \begin{aligned}
\left\langle\psi\left|{r_{12}^{-3}}\right| \psi\right\rangle
=-&\sum_{i=1}^{2} \left\langle\nabla_{i} \psi\left|{r_{12}^{-1}}{\ln r_{12}}\right| \nabla_{i} \psi\right\rangle \\
&+\left\langle\psi\left| 2\left(E_{\psi}-V\right) {\frac{\ln r_{12}}{r_{12}} }\right.\right.  \\
&\left.\left.+4 \pi(1+\gamma)\delta^{3}\left(r_{12}\right) \right|\psi\right\rangle  \,.\label{e10}
    \end{aligned}
\end{eqnarray}
where $r_{12}^{n}\ln r_{12}$ $(n=-2,-1,0,1)$ will be involved in integration. In addition to the above three terms, the expectation values of other operators appearing in Eqs.~(\ref{e5})-(\ref{e6}) will be calculated in the C-BSBF directly.

\subsection{Laplace's expansion of $r_{12}^{n}$ and $r_{12}^{n}\ln r_{12}$}

The integration of $r_{12}^{n}$ and $r_{12}^{n}\ln r_{12}$ are involved in the computation of Breit-Pauli operators and Araki-Sucher corrections. It is crucial to process this type of the integration in spherical coordinates, which requires separating their radial and angular dimensions. The generalization of Laplace's expansion to arbitrary powers and functions of $r_{12}$ given by Sack~\cite{sack64} is used to calculate the integration in which different powers of $r_{12}$ is involved. $r_{12}^n$ can be expanded in the form
\begin{eqnarray}
r_{12}^{n}=\sum\limits_{\ell=0}^{\infty} R_{n \ell}(r_{1}, r_{2}) P_{\ell}(\cos \theta_{12})\,,\label{e11}
\end{eqnarray} 
where the Legendre polynomials of $\cos \theta_{12}$ is expressed by using the identity as $P_{\ell}(\cos \theta_{12})=4\pi/(2\ell+1)\sum\limits_{m=-\ell}^{m=\ell}Y_{\ell m}^{\ast}(\hat{r}_1)Y_{\ell m}(\hat{r}_2)$, and the radial function $R_{n \ell}(r_{1}, r_{2})$ has been formulated by Sack~\cite{sack64} as following

\begin{eqnarray}
    \begin{aligned}
R_{n \ell}(r_{1}, r_{2})&=\dfrac{\left(-\frac{1}{2}n\right)_{\ell}}{\left(\frac{1}{2}\right)_{\ell}} r_{>}^{n}\left(\frac{r_{<}}{r_{>}}\right)^{\ell} \, \\
&\times\ _2F_1\left(l-\frac{1}{2}n, -\frac{1}{2}-\frac{1}{2}n; l+\frac{3}{2}; \frac{r_{<}^{2}}{r_{>}^{2}}\right)\,.\label{e12}
    \end{aligned}
\end{eqnarray}
In Eq.~(\ref{e12}), $r_{<}=\min(r_{1}, r_{2})$, $r_{>}=\max(r_{1}, r_{2})$, and the hypergeometric function has the form of

\begin{eqnarray}
_2F_1(\alpha, \beta; \gamma; x)=1+\sum\limits_1^{\infty}\dfrac{(\alpha)_s(\beta)_s}{(\gamma)_ss!}x^s\,,\label{e13}
\end{eqnarray}
where the Pochhammer symbol is defined as

\begin{eqnarray}
(\alpha)_s=\left\{
\begin{array}{ll}
1                                          &if\,s=0\\
\alpha(\alpha+1)\cdot\cdot\cdot(\alpha+s-1)&if\,s>0\\
\end{array}
\right.\,.
\end{eqnarray}

The hypergeometric function is finite series if either $\alpha$ or $\beta$ is zero or a negative integer, which implies that for all positive odd integer values of $n$, the series of $R_{n\ell}$ break off; and for $n=-1$, they consists of the leading term only. For positive even $n$, the summation is truncated to $\ell=\frac{n}{2}$, since the factor $(-\frac{1}{2}n)_{\ell}$ ensures that $R_{n\ell}$ vanishes when $\ell>\frac{n}{2}$. In addition, the individual functions $R_{n\ell}$ are divergent for $n\leq-2$, but they remain integrable as long as $n>-3$~\cite{yan96, frolov05}. Present calculations involve the integrations of $\langle\psi|r_{12}^{-2}|\psi\rangle$ and $\langle\psi|r_{12}^{-3}|\psi\rangle$. So giving appropriate radial expansions of $r_{12}^{-2}$ and $r_{12}^{-3}$ is important in the computation of radial and angular integrations. Substituting $n=-2\,,\ell=0$ and $n=-2\,,\ell=1$ separately into Eq.(\ref{e12}), and summation of the series, as a result the following specific expressions in terms of reverse hyperbolic tangent function $\tanh ^{-1}(x)$ are achieved,
\begin{eqnarray}
&&R_{-2, 0}\left(r_{1}, r_{2}\right)=\frac{\tanh ^{-1}(x)}{xr_{>}^{2}}\,,\\	
&&\begin{aligned}
    R_{-2, 1}\left(r_{1}, r_{2}\right)&= \frac{3}{2x^{2}r_{>}^{2}} \\
    \times &\left[\left(x^2+1\right) \tanh ^{-1}(x)-1\right]\,,
\end{aligned}
\end{eqnarray}			
where $x=r_{<}/r_{>}$; then the recurrence relation
\begin{eqnarray}
    \begin{aligned}			
\frac{r_{1}^{2}+r_{2}^{2}}{r_{1}r_{2}}R_{n,\ell}-\frac{\ell+2+\frac{1}{2}n}{\ell+\frac{3}{2}}R_{n,\ell+1}&\\
-\frac{\ell-1-\frac{1}{2}n}{\ell-\frac{1}{2}}&R_{n,\ell-1}=0\,,\label{e17}
\end{aligned}
\end{eqnarray}
can be used to calculate the radial functions for other values of $\ell$. For $n =-3$, the expansion coefficients of the hypergeometric functions are cancelled, and the hypergeometric functions are reduced to a series summation of $x^{n}$. the hypergeometric function can be expressed as analytic functions that is independent of $\ell$, correspondingly the radial expansion of $R_{-3, l}$ can be written as~\cite{lewis73}
\begin{eqnarray}
R_{-3, l}\left(r_{1}, r_{2}\right)=\frac{(2\ell+1)x^{\ell}}{\left(1-x^{2}\right)r_{>}^{3}}\,.
\end{eqnarray}

Next we will give the explicit formula for the product of $r_{12}$ with different powers and $\ln r_{12}$. Differentiation of Eq.~(\ref{e11}), the expansion for $r_{12}^{n}\ln r_{12}$ can be expressed as
\begin{eqnarray}
r_{12}^{n} \ln r_{12}=\sum_{\ell}R_{n \ln ,\ell}(r_{1},r_{2}) P_{\ell}\left(\cos \theta_{12}\right)\,,\label{e19} 	
\end{eqnarray}
where $R_{n\ln, \ell}(r_1, r_2)$ represents the radial function of $r_{12}^{n}\ln r_{12}$, and $R_{n\ln, \ell}(r_1, r_2)=\frac{\partial R_{n\ell}(r_{1},r_{2})}{\partial n}$.
Similarly, the following recurrence relation for $R_{n\ln, \ell}(r_1, r_2)$ can be derived by taking the derivative of Eq.~(\ref{e17}),

\begin{eqnarray}
\begin{aligned}	
    \frac{1}{2\ell+3}&R_{n,\ell+1}-\frac{1}{2\ell-1}R_{n,\ell-1}=\frac{r_{1}^{2}+r_{2}^{2}}{r_{1}r_{2}}R_{n \ln,\ell}\\
&-\frac{2\ell+4+n}{2\ell+3}R_{n\ln,\ell+1}-\frac{2\ell-2-n}{2\ell-1}R_{n\ln,\ell-1}\,.\label{e21} 
\end{aligned}
\end{eqnarray}
Then we can calculate the integration with the $r_{12}^{n}\ln r_{12}\,(n\geq-2)$ operator in the present paper. For example, for $n=-2\,,\ell=0$ and $n=-2\,,\ell=1$,

\begin{eqnarray}
&&R_{-2\ln, 0}=\frac{\tanh ^{-1}(x) \ln (r_{>}^{2}-r_{<}^{2})}{2 r_{>}^2 x}\,, \\
    &&\begin{aligned}		
R_{-2\ln, 1}=&\frac{3\left[\ln (r_{>}^{2}-r_{<}^{2})-1\right]}{4 r_{>}^2 x^2}\\
&\times \left[\left(x^2+1\right) \tanh ^{-1}(x)-x\right]\,, 
\end{aligned}
\end{eqnarray}
and the estimations of $R_{-2\ln, \ell}$ for other values of $\ell>1$ can be obtained according to the recurrence relation of Eq.~(\ref{e21}).

\section{results and discussions}

The C-BSBF on an exponential grid~\cite{bachau01} are generated using B-splines constrained to a spherical cavity. The cavity radius of $R_0=20$ a.u. is for the $1\,^1S$ state,  $R_0=40$ a.u. is for the $2\,^1S$ state, and $R_0=70$ a.u. is for both the $2\,^3S$ and $3\,^3S$ states. Yang et al.\cite{yang17} have implemented the correlated B-splines to calculate the helium atomic energy level and their non-relativistic ground state energy is $-$2.903 724 377 1(2) a.u.. A knot distribution optimization was performed for any individual states and present values of energies for the $1\,^1S$, $2\,^1S$, $2\,^3S$ and $3\,^3S$ states are listed in Table~\ref{t1}. The optimized result of $-$2.903 724 377 034 0(2) a.u. is obtained for the ground state, which has thirteen significant digits in agreement with Drake's. The $2\,^1S$, $2\,^3S$, and $3\,^3S$ states also reached fourteen significant digits in agreement with Drake.
%Optimizing the number of integration points for different integrands.

%������۲���ע��Ҫ˵��һ��Ŀǰ����\Delta_1^2\Delta_2^2�����ѣ��Լ�Hylleraas����������������շ�껵Ľ���Hylleraas�������ļ����ǽ�����
\begin{table}[!htbp]
    \caption{\label{t1} Energies for the $1\,^1S$, $2\,^1S$, $2\,^3S$ and $3\,^3S$ states of helium.}
    \begin{ruledtabular}
    \begin{tabular}{clll}
    \multicolumn{1}{c}{State} &\multicolumn{1}{c}{This work}          &\multicolumn{1}{c}{Ref.\cite{drake06}}\\ \hline
    $1^{1}S$                  &$-$2.903 724 377 034 0(2)              &$-$2.903 724 377 034 119 5 \\
    $2^{1}S$                  &$-$2.145 974 046 054 4(2)              &$-$2.145 974 046 054 419(6) \\
    $2^{3}S$                  &$-$2.175 229 378 236 7(2)              &$-$2.175 229 378 236 791 30 \\
    $3^{3}S$                  &$-$2.068 689 067 472 4(2)              &$-$2.068 689 067 472 457 19
    \end{tabular}
    \end{ruledtabular}
    \end{table}

\begin{table*}[!htbp]
    \begin{threeparttable}
    \caption{\label{ta} The expectation values of other operators needed for evaluating the relativistic kinetic terms for the $1\,^1S$, $2\,^1S$, $2\,^3S$, and $3\,^3S$ states of helium. }
    \begin{ruledtabular}
    \begin{tabular}{cllll}
     \multicolumn{1}{c}{$Operater$}                &\multicolumn{1}{c}{$1^{1}S$}    &\multicolumn{1}{c}{$2^{1}S$}         &\multicolumn{1}{c}{$2^{3}S$}          &\multicolumn{1}{c}{$3^{3}S$}                   \\ \hline
     \multirow{3}{*}{$\langle1/r_{1}\rangle$}      &1.688 316 800 717 1(2)          &1.135 407 686 126 1(2)               &1.154 664 152 972 0(1)                &1.063 674 075 760 7(2)                         \\
                                                   &1.688 316 800 717\tnote{a}      &1.135 407 686 125 609(6)\tnote{b}    &1.154 664 152 972 107 60(20)\tnote{b} &1.063 674 075 760 76(10)\tnote{b}              \\
                                                   &1.688 316 800 635\tnote{c}      &1.135 407 686\tnote{c}               &1.154 664 152\tnote{c}                &1.063 674 075 7\tnote{c}                       \\
    \multicolumn{5}{c}{}                                                                                                                                                                                         \\ 
    \multirow{2}{*}{$\langle1/r_{1}^{2}\rangle$}   &6.017 408 867 0(3)              &4.146 939 019 80(6)                  &4.170 445 551 31(2)                   &4.042 948 747 4(3)                             \\
                                                   &6.017 408 867 0(1)\tnote{a}     &4.146 939 019 0(12)\tnote{b}         &4.170 445 551 336 2(4)\tnote{b}       &4.042 948 747 477(4)\tnote{b}                  \\
    \multicolumn{5}{c}{}                                                                                                                                                                                         \\  
    \multirow{2}{*}{$\langle1/r_{1}r_{2}\rangle$}  &2.708 655 474 480(4)            &0.561 861 467 461(2)                 &0.560 729 635 682 9(3)                &0.240 684 804 629 3(2)                         \\
                                                   &2.708 655 474 480\tnote{a}      &0.561 861 467 459 6(7)\tnote{b}      &0.560 729 635 682 926 40(20)\tnote{b} &0.240 684 804 629 353(11)\tnote{b}             \\
    \multicolumn{5}{c}{}                                                                                                                                                                                         \\ 
    \multirow{3}{*}{$\langle1/r_{12}\rangle$}      &0.945 818 448 799 95(5)         &0.249 682 652 394 3(6)               &0.268 197 855 414 82(5)               &0.117 318 168 097 65(4)                        \\
                                                   &0.945 818 448 800\tnote{a}      &0.249 682 652 393 566 7(19)\tnote{b} &0.268 197 855 414 847 80(20)\tnote{b} &0.117 318 168 097 636(6)\tnote{b}              \\
                                                   &0.945 818 448 705 9\tnote{c}    &0.249 682 652 3\tnote{c}             &0.268 197 855 3\tnote{c}              &0.117 318 168 0\tnote{c}                       \\
    \multicolumn{5}{c}{}                                                                                                                                                                                         \\
    \multirow{2}{*}{$\langle1/r_{1}r_{12}\rangle$} &1.920 943 921 900 0(5)          &0.340 633 845 861 2(8)               &0.322 696 221 719 8(2)                &0.131 426 560 051 19(5)                        \\
                                                   &1.920 943 921 900\tnote{a}      &0.340 633 845 861 0(19)\tnote{b}     &0.322 696 221 719 854 32(8)\tnote{b}  &0.131 426 560 051 184(5)\tnote{b}              \\
    \end{tabular}
    \end{ruledtabular}
    \begin{tablenotes}
        \item[a] Drake \cite{drake06}.
        \item[b] Drake \cite{drakenote}.
        \item[c] Yu et al. \cite{yu22}.
        \end{tablenotes}
    \end{threeparttable}
    \end{table*}

    It can be seen from Eq.~(\ref{P^{4}}) that the computation of $\langle p_1^4\rangle$ involves many operators, which are classified into two categories for dealing with. One type is the general operators that are relatively simple to compute, including $1/r_{1}$, $1/r_{1}^{2}$, $1/r_{1}r_{2}$, $1/r_{12}$ and $1/r_{1}r_{12}$. We give the final convergence values directly in Table~\ref{ta}, and there are at least ten significant digits of our results that are consistent with Drake’s. This also demonstrates the high accuracy of the wave function obtained for the C-BSBF.

%================================================================================================
\begin{table*}[!htbp]
\caption{\label{t2} Convergence of the relativistic kinetic terms for the $1\,^1S$, $2\,^1S$, $2\,^3S$ and $3\,^3S$ states of helium as the number of B-splines $N$ increased. The expectation values of $1/r_{12}^2$ and $\nabla_1^2\nabla_2^2$ are also listed in the second and third columns. The partial wave is $\ell_{max}=4$.  }
\begin{ruledtabular}
\begin{tabular}{clll}
\multicolumn{1}{c}{$N$} &\multicolumn{1}{c}{$\langle1/r_{12}^2\rangle$} &\multicolumn{1}{c}{$\langle\nabla_1^2\nabla_2^2\rangle$} &\multicolumn{1}{c}{$\langle p_1^4\rangle$} \\
\hline
\multicolumn{4}{c}{$1\,^1S$}\\
50                     &1.464 770 923 579             &7.133 709 835                &54.088 067 177              \\
60                     &1.464 770 923 463             &7.133 709 771                &54.088 067 242              \\
70                     &1.464 770 923 406             &7.133 709 763                &54.088 067 251              \\
Extrap.                &1.464 770 923 3(5)            &7.133 709 7(2)               &54.088 067 2(2)             \\
Ref.~\cite{patkos17}   &1.464 771                     &7.133 710                    &                            \\
Ref.~\cite{drake06}    &1.464 770 923 350(1)          &                             &54.088 067 230(2)           \\
\multicolumn{4}{c}{}\\
\multicolumn{4}{c}{$2\,^1S$}\\
50                     &0.143 724 814 027             &1.428 212 689 1              &41.118 675 563 8            \\
60                     &0.143 724 814 013             &1.428 212 706 4              &41.118 675 546 0            \\
70                     &0.143 724 814 008             &1.428 212 705 8              &41.118 675 546 6            \\
Extrap.                &0.143 724 814 00(5)           &1.428 212 70(4)              &41.118 675 54(4)            \\
Ref.~\cite{patkos17}   &0.143 725                     &1.428 213                    &                            \\
Ref.~\cite{drakenote}  &0.143 724 814 00(7)           &                             &41.118 675 544(19)          \\
\multicolumn{4}{c}{}\\
\multicolumn{4}{c}{$2\,^3S$}\\
50                     &0.088 906 004 870             &0.488 197 568 41             &41.835 540 798 28           \\
60                     &0.088 906 004 913             &0.488 197 569 31             &41.835 540 797 46           \\
70                     &0.088 906 004 921             &0.488 197 569 91             &41.835 540 796 85           \\
Extrap.                &0.088 906 004 9(2)            &0.488 197 570(4)             &41.835 540 796(4)           \\
Ref.\cite{patkos16}    &0.088 906                     &0.488 198                    &                            \\
Ref.\cite{drakenote}   &0.088 906 004 932 625(5)      &                             &41.835 540 797 348(6)       \\
\multicolumn{4}{c}{}\\
\multicolumn{4}{c}{$3\,^3S$}\\
50                     &0.023 097 669 645             &0.329 220 596 46            &40.475 439 870 27            \\
60                     &0.023 097 669 653             &0.329 220 596 68            &40.475 439 868 42            \\
70                     &0.023 097 669 655             &0.329 220 596 89            &40.475 439 868 25            \\
Extrap.                &0.023 097 669 65(3)           &0.329 220 597(2)            &40.475 439 868(5)            \\
Ref.\cite{drakenote}   &0.023 097 669 656 893(13)     &                            &40.475 439 868 127 2(3)      \\
\end{tabular}
\end{ruledtabular}
\end{table*}
The other type is the operators $1/r_{12}^2$ and $\nabla_1^2\nabla_2^2$ that are more difficult to calculate. The numerical results of $\langle 1/r_{12}^2\rangle$, $\langle\nabla_1^2\nabla_2^2\rangle$ and $\langle p_1^4\rangle$ as the number of B-splines $N$ increased are given in the last three columns of Table~\ref{t2}. Good convergent values of $\langle1/r_{12}^2\rangle$ under the C-BSBF are achieved with the global operator method. For the ground state, the present result of 1.464 770 923 3(5) is obtained, which has eleven significant figures and agrees well with reference values with the explicitly correlated exponential basis~\cite{patkos17} and the Hylleraas basis~\cite{drake06}. Our expectation values of $1/r_{12}^2$ for the $2\,^1S$, $2\,^3S$ and $3\,^3S$ states of the helium atom at least have eight convergent digits, which are all in good agreement with results in available literatures~\cite{patkos16, patkos17, drakenote}. For the $\langle\nabla_1^2\nabla_2^2\rangle$ operator, no suitable treatment could be found to make it converge faster, for which the direct calculation method was used. Therefore, the convergent accuracy of $\langle\nabla_1^2\nabla_2^2\rangle$ is relatively lower, which is also the main reason to limit the numerical precision of $\langle p_1^4\rangle$. The present result of $\langle p_1^4\rangle$ for the $1\,^1S$ state from the C-BSBF has nine digits, consistent with Drake's Hylleraas results~\cite{drakenote,drake06}. Present numerical convergence for the triplet states are better than for the singlet states by one to two significant figures, and our values are both good agreement with Hylleraas results~\cite{drakenote}. 

We also calculated $\langle 1/r_{12}^2\rangle$ and $\langle\nabla_1^2\nabla_2^2\rangle$ using the traditional B-spline basis set, and results for the ground state are 1.463 697 and 7.079, respectively. Since these singularity operators only have one to three significant digits, which are difficult to use in high-precision calculations at the atomic energy level. It is convenient to find that the primary explanation for this is that the traditional B-spline basis set makes it difficult to describe the local properties of the wave function with high accuracy without including the electron correlation effect.

%================================================================================================
\begin{table*}[!htbp]
\caption{\label{t3} The expectation values of $\delta^3 (r_{1})$, $\delta^3 (r_{12})$, $H_2$ and $1/r_{12}^{3}$ for the $1\,^1S$, $2\,^1S$, $2\,^3S$ and $3\,^3S$ states of helium. Comparisons with results obtained in available literatures are also made. The partial wave is $\ell_{max}=4$. }
\begin{ruledtabular}
\begin{tabular}{cllll}
\multicolumn{1}{c}{$N$} &\multicolumn{1}{c}{$\langle\delta^3(r_{1})\rangle$} &\multicolumn{1}{c}{$\langle\delta^3(r_{12})\rangle$} &\multicolumn{1}{c}{$\langle H_2\rangle$}
&\multicolumn{1}{c}{$\langle1/r_{12}^{3}\rangle$}\\
\hline
\multicolumn{5}{c}{$1^{1}S$} \\
50                       &1.810 429 325 97         &0.106 345 370 649 3         &$-$0.139 094 671 8            &0.989 271 57            \\
60                       &1.810 429 323 14         &0.106 345 370 658 3         &$-$0.139 094 675 1            &0.989 271 98            \\
70                       &1.810 429 321 51         &0.106 345 370 646 3         &$-$0.139 094 677 3            &0.989 272 26            \\
Extrap.                  &1.810 429 32(2)          &0.106 345 370 66(4)         &$-$0.139 094 67(2)            &0.989 272(2)            \\
Ref.\cite{yu22}          &1.810 429 318 371 521 8  &0.106 346 068               &                              &                        \\
Ref.\cite{drakenote}     &1.810 429 318 499 0(6)   &0.106 345 370 636 3(12)     &$-$0.139 094 690 539 20(20)   &0.989 273 544 768(13)   \\
Ref.\cite{pachucki05}    &                         &                            &                              &0.989 273 5             \\
Ref.\cite{pachucki00}    &                         &                            &                              &0.989 272 4(13)         \\
\multicolumn{5}{c}{} \\
\multicolumn{5}{c}{$2^{1}S$} \\
50                       &1.309 460 780 907        &0.008 648 433 612 1         &$-$0.009 253 044 67           &0.067 946 402           \\
60                       &1.309 460 780 719        &0.008 648 433 588 4         &$-$0.009 253 044 78           &0.067 946 439           \\
70                       &1.309 460 780 607        &0.008 648 433 587 3         &$-$0.009 253 044 97           &0.067 946 465           \\
Extrap.                  &1.309 460 780 5(8)       &0.008 648 433 58(5)         &$-$0.009 253 045(2)           &0.067 946 4(2)          \\
Ref.\cite{yu22}          &1.309 460 780 3          &0.008 648 6                 &                              &                        \\
Ref.\cite{drakenote}     &1.309 460 780 1(4)       &0.008 648 433 6(14)         &$-$0.009 253 046 05(4)        &                        \\
Ref.\cite{drake92}       &                         &                            &                              &0.067 946 32            \\
\multicolumn{5}{c}{} \\
\multicolumn{5}{c}{$2^{3}S$} \\
50                       &1.320 355 082 933 78     &                            &$-$0.001 628 430 082 9        &0.038 861 479 8         \\
60                       &1.320 355 082 931 58     &                            &$-$0.001 628 430 067 4        &0.038 861 479 6         \\
70                       &1.320 355 082 931 10     &                            &$-$0.001.628 430 064 8        &0.038 861 481 0         \\
Extrap.                  &1.320 355 082 930(6)     &                            &$-$0.001 628 430 06(4)        &0.038 861 46(3)         \\
Ref.\cite{yu22}          &1.320 355 082 9          &                            &                              &                        \\  
Ref.\cite{drakenote}     &1.320 355 082 934 92(9)  &                            &$-$0.001 628 430 061 553(3)   &                        \\
Ref.\cite{drake92}       &                         &                            &                              &0.038 861 485 631 95    \\
\multicolumn{5}{c}{} \\
\multicolumn{5}{c}{$3^{3}S$} \\
50                       &1.285 060 253 969 23     &                            &$-$0.000 504 504 232 33       &0.008 922 569 5         \\
60                       &1.285 060 253 936 06     &                            &$-$0.000 504 504 228 95       &0.008 922 569 6         \\
70                       &1.285 060 253 938 13     &                            &$-$0.000 504 504 228 33       &0.008 922 569 9         \\
Extrap.                  &1.285 060 253 93(7)      &                            &$-$0.000 504 504 228(9)       &0.008 922 57(2)         \\
Ref.\cite{yu22}          &1.285 060 253 9          &                            &                              &                        \\
Ref.\cite{drakenote}     &1.285 060 253 932 1(13)  &                            &$-$0.000 504 504 227 201(9)   &                        \\
\end{tabular}
\end{ruledtabular}
\end{table*}
The expectation values of other three components from $H_{BP}$ and the singular electron-electron $\langle 1/r_{12}^{3}\rangle$ from the leading QED corrections are shown in Table~\ref{t3}. The expectation values of $\delta^3(r_{12})$ for the triplet states equal zero, so they are not listed in Table~\ref{t3}. Yu et al.~\cite{yu22} employed the same C-BSBF to give numerical results of $\delta^3(r_1)$ by direct calculation when the power of $r_{12}$ is $c=5$, which are also shown in Table~\ref{t3}. The direct calculation of $\delta^3(r_1)$ is highly dependent on the origin value of the wave function, and the global operator method can be used to further improve the calculation accuracy. The result of the $\delta^3(r_{1})$ of the ground state using the global operator method is 1.810 429 32(2), one can see that numerical accuracy of the $\delta^3(r_{1})$ can reach a precision of eight to twelve signiﬁcant digits. It can be seen that our computational accuracy with $c=1$ is completely comparable to theirs~\cite{yu22}, with the except for the ground state with relatively sensitive electron correlations. They also tried to improve the direct calculation accuracy of $\delta^3(r_{1})$ by increasing the power of $r_{12}$, but the global operator method is still necessary to effectively improve the numerical convergence. For example, our result of $\langle\delta^3(r_{12})\rangle$ for the $1\,^1S$ state is 0.106 345 370 66(4), that is more accurate than 0.106 346 068 of Yu et al.~\cite{yu22} by five orders of magnitude and is well consistent with Drake's Hylleraas value of 0.106 345 370 636 3(12)~\cite{drakenote} as well. Present results for the retardation term $H_2$ have at least seven convergent figures and agree with Drake's~\cite{drakenote}. The expectation of singular electron-electron $\langle 1/r_{12}^{3}\rangle$ are computed with the global operator method by the C-BSBF and confronted with previous results obtained from different basis functions as well. Present the C-BSBF result of 0.989 272(2) with an accuracy of five decimals is achieved for the ground state, which is comparable to results of 0.989 273 5 and 0.989 272 4(13) with explicitly correlated Gaussian (ECG) functions~\cite{pachucki05} and exponential basis functions~\cite{pachucki00}, respectively, in numerical precision. Employed Hylleraas basis and exponential basis respectively, Drake~\cite{drakenote} improved reference values with three additional exact digits. Our result for the ground state is expected to recover more figures of Drake's result if adopting higher power of $r_{12}$. For the $2\,^1S$ and $2\,^3S$ states, our values are in agreement with previous values obtained by Hylleraas basis and exponential basis~\cite{drake88}. There are five convergent figures in our result $\langle1/r_{12}^{3}\rangle$=0.008 922 57(2) for the $3\,^3S$ state. 

The singular electron-electron $\langle 1/r_{12}^{3}\rangle$ expectation value is also computed using the traditional B-spline basis set, and the ground state result is 1.197($N=70$, $\ell_{max}=4$). It can be seen that the traditional B-spline is entirely inaccurate in calculating $\langle 1/r_{12}^{3}\rangle$, and this type of operator for divergence require a more accurate description of the local properties of the wave function \cite{pachucki05} than $1/r_{12}^2$ and $\nabla_1^2\nabla_2^2$. As a result, the B-spline basis set containing electron correlation is essential.

The final relativistic corrections are presented in the top half of Table~\ref{t4}. Comparisons are made with results obtained using the explicitly correlated exponential basis~\cite{yerokhin10} and Hylleraas basis~\cite{drakenote}. Our relativistic corrections results are completely consistent with most precise previous calculations~\cite{yerokhin10, drakenote} and can reach eight to ten significant figures. The leading QED corrections for the $S$ states to the energy level are summarized in the bottom half of Table~\ref{t4}, which used the Bethe logarithm values obtained from B-splines \cite{yang19,Zhang20}, and Korobov's Bethe logarithm values~\cite{korobov19} as a benchmark, respectively. It can be seen that our calculated results are in good agreement with the significant figures listed by Yerokhin et al.\cite{yerokhin10}, where the results of the singlet state calculated using Korobov's Bethe logarithm values are almost identical to the results from B-splines, which are mainly explained by the relatively low accuracy of $\delta^{3}(r_{1})$ and $1/r_{12}^{3}$, and the improved accuracy of the triplet state is the result of the limited accuracy of Bethe logarithm. The overall computational accuracy of the leading QED correction is determined mainly by the contribution of the Araki-Sucher term and $\delta^{3}(r_{1})$ for the ground state, by the contribution of the Bethe logarithms for other states. It can be seen that the leading QED corrections results can reach at least seven significant digits, which already reaches the accuracy level of the contribution of the leading relativistic correction in this work. In addition, the numerical accuracy of the singlet is expected to improve with increasing power $c$ of $r_{12}$ in basis function.

\begin{table*}[!htbp]
    \caption{\label{t4} The leading relativistic and QED corrections, $\delta E_{rel}$ and $\delta E_{QED}$ for the $1\,^1S$, $2\,^1S$, $2\,^3S$ and $3\,^3S$ states of helium. The corresponding comparison data given in available literatures are also listed.}
    \begin{ruledtabular}
    \begin{tabular}{cllll}
                                                                   &\multicolumn{1}{c}{$1^{1}S$}   &\multicolumn{1}{c}{$2^{1}S$}   &\multicolumn{1}{c}{$2^{3}S$}      &\multicolumn{1}{c}{$3^{3}S$}           \\
    \hline           
    \multicolumn{5}{c}{the leading relativistic correction} \\           
    $\delta E_{rel}/\alpha^2$                                      &$-$1.951 754 7(2)              &$-$2.034 167 33(2)             &$-$2.164 477 971(2)                &$-$2.045 092 764(2)      \\
    Ref.\cite{drakenote}                                           &$-$1.951 754 767               &$-$2.034 167 342               &$-$2.164 477 972                  &$-$2.045 092 764        \\               
    \multicolumn{5}{c}{}    \\                
    \multicolumn{5}{c}{the leading QED correction} \\                
    $\delta E_{QED}/\alpha^3$(BL with B-splines)       &57.288 165(2)                  &42.523 605 2(2)                &43.010 017(2)                     &41.839 303 4(7)         \\
    $\delta E_{QED}/\alpha^3$(BL from Korobov)        &57.288 165(1)                  &42.523 605 10(8)                &43.010 017 06(2)                  &41.839 301 459(9)         \\
    Ref.\cite{yerokhin10}                             &57.288 165 2                   &42.523 605 1                   &43.010 016 8                      &                        \\
    \end{tabular}
    \end{ruledtabular}
\end{table*}

\section{Summary and outlook}
In this work, we have calculated the leading relativistic and QED corrections of the energy levels of the helium atom using the C-BSBF. The expectation values of the relativistic kinetic term $p_{1}^{4}$, contact potential $\delta^{3}(r_{1})$, $\delta^{3}(r_{12})$ and Araki-Sucher correction $\langle 1/r_{12}^{3} \rangle$, which are more difficult to calculate directly, were treated by a global operator method to improve their numerical convergence, and the two-electron distance function is also introduced to deal with the Laplace expansion method proposed by Sack~\cite{sack64}. Together with the high-precision calculation of the Bethe logarithms~\cite{yang19}, the C-BSBF is able to achieve the high-precision calculation of the leading relativistic and QED corrections for the energy levels of the helium atom. It is emphasized that the correlated factor $r_{12}$ in the C-BSBF is crucial to calculate $p_{1}^{4}$, $\delta^{3}(r_{12})$ and $\langle 1/r_{12}^{3} \rangle$, without this factor, these operators have a very slow convergence. The C-BSBF can provide stable numerical convergence based on its approximate linear independence and sufficient consideration of the electronic correlation. It can be seen from Table~\ref{t5} that the C-BSBF can determine the accuracy of the $2^{3}S-2^{1}S$ transition frequency (up to $m\alpha^{5}$-order correction) to the kHz level, which is consistent with the results of Pachucki et al., reaching a level similar to the latest experiment~\cite{rengelink18}. A further improvement in our results is expected, if adopting higher power of $r_{12}$. The calculations are carried out applying double precision, no multi-precision is needed. This method provides a new approach to the calculation of atom energy levels. 

\begin{table}[!htbp]
    \caption{\label{t5} The $2^{3}S-2^{1}S$ transition frequency for the helium atom along the leading relativistic and QED corrections, in KHz.}
    \begin{ruledtabular}
    \begin{tabular}{crr}
                  &\multicolumn{1}{c}{$\Delta E(2^{3}S-2^{1}S)$}   &\multicolumn{1}{c}{Ref.\cite{pachucki17}} \\ \hline
    NR            & 192 490 838 748(2)        & 192 490 838 756 \\
    $m\alpha^4$    &      45 657 862(8)        &      45 657 859   \\
    $m\alpha^5$    &    $-$1 243 669(6)        &    $-$1 243 671   \\ \hline
    Expt.~\cite{rengelink18} &\multicolumn{2}{c}{192 510 702 148.72(20)} \\
    \end{tabular}
    \end{ruledtabular}
\end{table}

Recently, Mitroy and Tang suggested testing the QED theory using tune-out wavelength, which opens a new way to test fundamental atomic structure theory~\cite{Mitroy13}. The 413~nm tune-out wavelengths for the helium atom $2^{3}S_1$ state discrepancies in the latest experiments by Baldwin's team and theoretical values by Drake based on the Hylleraas basis set with the NRQED method~\cite{Henson2022}, in which the calculation only estimates the electric-ﬁeld dependence of the Bethe logarithm~\cite{grzegorz04}. The precision of the experiment is expected to improve further, and the QED theory will be tested at higher precision. The ab-initio calculation of the electric field dependence of the Bethe logarithm is important for further improving the theoretical prediction accuracy of the 413~nm tune-out wavelength to a level of ppb. The successful application of the C-BSBF in singular operator calculations in this work suggests that the C-BSBF is expected to be used to calculate the electric field dependence of Bethe logarithms to improve the theoretical calculation accuracy of the 413~nm tune-out wavelength. In addition to the C-BSBF is also expected to be extended to the second-order perturbation of the Breit–Pauli operators~\cite{pachucki06} and relativistic corrections to the Bethe logarithm~\cite{Yerokhin18} of helium atom in the future.

\section{ackonwledgement}
This work is supported by the National Natural Science Foundation of China under Grants No. 12274423 and No. 12274417, by the Chinese Academy of Sciences Project for Young Scientists in Basic Research under Grant No. YSBR-055.

%\bibliographystyle{apsrev_title.bst}
%\bibliography{positron.bib}

\end{document}